\documentclass[letterpaper,10pt,conference]{ieeeconf}
\IEEEoverridecommandlockouts
\usepackage[T1]{fontenc}
\usepackage{amsmath,amsfonts,bm}
\usepackage{cite}
\usepackage{graphicx}
\usepackage{array}
\usepackage{booktabs}
\usepackage{makecell}
\usepackage{multirow}
\usepackage{longtable}
\usepackage{tabularx}
\usepackage{balance}

\usepackage{algorithm}
\usepackage{algpseudocode}
\usepackage{placeins} 

\usepackage{textcomp}
\usepackage{xurl}
\usepackage{scalerel}
\usepackage{xhfill}
\usepackage{stfloats}
\usepackage{xspace}
\usepackage{xcolor}

\usepackage[caption=false]{subfig}
\usepackage[font=normalsize]{caption}
\def\p(#1|#2){p\left(#1\,|\,#2\right)}

\newtheorem{rem}{Remark}

\let\originalleft\left
\let\originalright\right
\renewcommand{\left}{\mathopen{}\mathclose\bgroup\originalleft}
\renewcommand{\right}{\aftergroup\egroup\originalright}

\begin{document}

\title{System Identification of Lithium-Ion Battery Equivalent Circuit Models Using Ensemble Kalman Inversion}
\IEEEoverridecommandlockouts
\author{
    Farzaneh Barat\textsuperscript{1}, Sara Wilson\textsuperscript{1}, Huijeong Kim\textsuperscript{2}, and Huazhen Fang\textsuperscript{3}%
    \thanks{\textsuperscript{1}F. Barat and S. Wilson are with the Department of Mechanical Engineering, University of Kansas, Lawrence, KS 66045, USA. {\tt\small \{barat, sewilson\}@ku.edu}}%
    \thanks{\textsuperscript{2}H. Kim is with the Department of Civil, Environmental \& Architectural Engineering, University of Kansas, Lawrence, KS 66045, USA. {\tt\small hjkim@ku.edu}}%
    \thanks{\textsuperscript{3}H. Fang is with the Department of Mechanical Engineering, Michigan State University, East Lansing, MI 48824, USA. {\tt\small hfang@msu.edu}}%
}

\maketitle

\begin{abstract}
System identification remains an intriguing challenge for lithium-ion batteries, as many models are nonlinear, exhibit multi-physics coupling, and involve a large number of parameters. In this paper, we address this challenge using the ensemble Kalman inversion (EnKI) method for battery system identification. EnKI performs maximum a posteriori parameter estimation through successive local Gaussian approximations, enabling an iterative and incremental search for unknown parameters. The search combines Monte Carlo sampling with Kalman-type updates to evolve an ensemble of samples, thereby offering empirical stability and the ability to handle strongly nonlinear models. We validate the proposed approach on two equivalent circuit models with coupled electro-thermal dynamics, through both simulation and experiments. The results demonstrate that the proposed approach achieves accurate parameter estimation with rapid iterative convergence, and it shows strong potential for application to other battery models.
\end{abstract}

\section{Introduction}

Lithium-ion batteries (LiBs) are transforming the sectors of transportation, grid infrastructure, and renewable energy as the arguably foremost energy storage technology of our time~\cite{NGOY:2025:RSE, KAPOOR:2025:MSE}. Essential to their operation are battery management systems, which perform condition monitoring,   thermal regulation, charging control, and fault detection, among other functions. The effectiveness of these systems relies  on the availability of accurate battery models, making dynamic modeling and model identification fundamentally important~\cite{plett:2015:bms}.

Equivalent circuit models (ECMs) have proven to be one of the most important classes of models for LiBs, thanks to their simple structures  and efficient computation~\cite{HU:2012:PS}. They use circuit analogs composed of voltage sources, resistors, and capacitors to approximate the dynamic behaviors of batteries. Many widely used ECMs originate from phenomenological modeling, which emphasizes reproducing the observed response during charging and discharging. A representative example is the Rint model, which consists of an open-circuit voltage (OCV) source in series with a single resistor~\cite{He:2011:E}. The OCV is expressed as a function of the state of charge (SoC), while the resistor accounts for the instantaneous voltage drop or recovery under current loading or unloading. The Thevenin model extends this structure by including resistor–capacitor (RC) pairs to capture transient responses and polarization effects~\cite{MOUSAVIG:2014:RSE}. Recent research has advanced ECM development by drawing inspiration from electrochemical models to enhance accuracy while retaining computational efficiency. In this vein, a leading example is the nonlinear double-capacitor (NDC) model, which employs an RC circuit to mimic lithium-ion diffusion within the electrode and couples it with a Thevenin-type circuit to represent terminal voltage dynamics~\cite{Tian:2021:TCS}. This idea has been expanded further: an extended NDC model is proposed in~\cite{DeOliveira:Heliyon:2024}, and the BattX model combines multiple circuit approximations of diffusion in both the electrode and  electrolyte, thereby enabling accurate predictions from low to high C-rates~\cite{BIJU:2023:AE}. Other work has approximated pseudo-2D electrochemical models using transmission-line circuit structures~\cite{GENG:2021:EA}. 
Beyond electrical behaviors, ECMs have gained wide use in thermal modeling of LiBs. For cylindrical cells, thermal circuit models have been developed that lump the spatial temperature distribution into core and surface nodes~\cite{LIN:2014:PS}. For pouch cells, more sophisticated  thermal circuit networks can offer fine spatial resolution in temperature variation~\cite{YUAN:2025:ATE}.

For nearly all models, system identification is as important as structure design, and ECMs are no exception. System identification seeks to extract a model’s unknown parameters from experimental data~\cite{TIAN:2020:ES}, and it is essential for ensuring the predictive accuracy and practical applicability of ECMs in battery engineering. The literature includes two main approaches as reviewed below.  

{\em Experiment-based model calibration.} This approach designs and uses  charging and discharging experiments to approximately determine model parameters. For example, trickle-current tests can be used to identify the SoC-OCV relationship, while pulse tests allow calibration of RC parameters by fitting the transient responses~\cite{Chen:2006:IEEE, Zhang:2020:en}. This method is  effective for identifying simple models such as the Rint model or the Thevenin model with one or two RC pairs. To account for temperature dependence, one can conduct pulse tests  under different thermal conditions to quantify how internal resistance varies with temperature~\cite{LUDWIG:2021:PS}. For more sophisticated models such as the BattX model, it is possible to design  multi-pronged experiments to excite different dynamic processes---such as lithium-ion diffusion in the electrode or electrolyte---so as to determine the corresponding circuit parameters. In general, experiment-based calibration is straightforward to understand and implement. However, its accuracy is often limited, and estimation uncertainty may accumulate when multiple tests are used to calibrate different groups of parameters.

{\em Optimization-based model identification.} Parameter estimation is also a data-fitting problem from an optimization perspective---it attempts to minimize the discrepancy between model predictions and measurement data subject to the model structure, often leading to nonlinear least-squares formulations~\cite{FENG:2015:PS, Sitterly:2011:IEEE}. One can set up such problems to identify ECMs, which can be solved easily when the excitation input is a constant current, since the model’s response admits an explicit parameterization. However, constant-current excitations provide limited information about the unknown parameters and thus constrain estimation accuracy. Richer excitation signals, like variable charging/discharging profiles, generate more informative data and thereby improve parameter estimation~\cite{Samieian:2022:B}. To leverage such data, the study in~\cite{Tian:2021:TCS} exploits the prediction-error method to identify the NDC model by treating it as a Wiener-type structure. This method yet applies to only ECMs with linear state dynamics. For nonlinear ECMs, parameter estimation becomes considerably more challenging: the resulting optimization problems are nonlinear, nonconvex, and naturally high-dimensional due to latent state variables. 
Gradient-based optimization methods often struggle with these challenges, while evolutionary algorithms--such as genetic algorithms~\cite{MALIK:2014:PS}, particle swarm optimization~\cite{YU:2017:IEEE}, and differential evolution~\cite{YANG:2014:PS}--have proven useful. Bayesian optimization offers another way by constructing a probabilistic surrogate of the objective function and using the surrogate to guide the search for optima~\cite{Tu:2024:ACC}. Both evolutionary and Bayesian optimization methods are gradient-free and possess global search capabilities, but they are computationally intensive and their practical performance depends  on initialization, population size, and iteration budgets, among others.

To date, system identification for nonlinear ECMs remains an open challenge. It requires solving high-dimensional, nonlinear, nonconvex optimization problems, making effective and efficient identification nontrivial. In this paper, we propose an alternative approach based on ensemble Kalman inversion (EnKI) as the key contribution. EnKI addresses maximum a posteriori (MAP) estimation from a probabilistic perspective, which can be recast as a weighted nonlinear least-squares problem~\cite{Iglesias:2013:IP, Kovachki:2019:IP}. It employs ensembles of samples to approximate concerned probability distributions and evolves these ensembles iteratively until convergence. By design, EnKI combines Monte Carlo sampling with a Kalman-type update, yielding  gradient-free and empirically stable search~\cite{Iglesias:2018:IP}. With these features, EnKI offers several attractive merits for parameter estimation, including the ability to handle strongly nonlinear models, to accommodate a relatively large number of unknown parameters, and to naturally account for noise in both the dynamic and measurement processes.

Centering on this key contribution, we consider two nonlinear ECMs: a temperature-dependent Thevenin model and a temperature-dependent NDC model. For these models, we present a systematic development of the EnKI-based system identification approach. In doing so, we introduce a new perspective on the underlying logic of the EnKI method, which, to the best of our knowledge, has not been reported in the literature. We then validate the approach using both simulation and experimental data, demonstrating its effectiveness in extracting parameters for the considered models. Beyond the considered ECMs, we emphasize that the proposed approach is  applicable to other ECMs and electrochemical models, and more broadly, to  generic state-space dynamic models in other application domains.

The remainder of this paper is organized as follows. Section~\ref{sec:ECM} reviews the considered ECMs, and Section~\ref{sec:SysID-EnKI} develops the EnKI-based system identification approach for them. Section~\ref{sec:Numerical_Simulation} evaluates the effectiveness of the proposed approach using both simulation and experiments. Finally, Section~\ref{sec:Conclusion} concludes the paper. 

\section{Overview of Considered ECMs}
\label{sec:ECM}
This section presents two ECMs that integrate electrical and thermal dynamics: a temperature-dependent Thevenin model (TheveninT) and a temperature-dependent NDC model (NDCT). Both models are formulated within a unified nonlinear state-space framework to facilitate system identification.
 
\begin{figure}[b]
\centering
\includegraphics[width=0.48\textwidth]{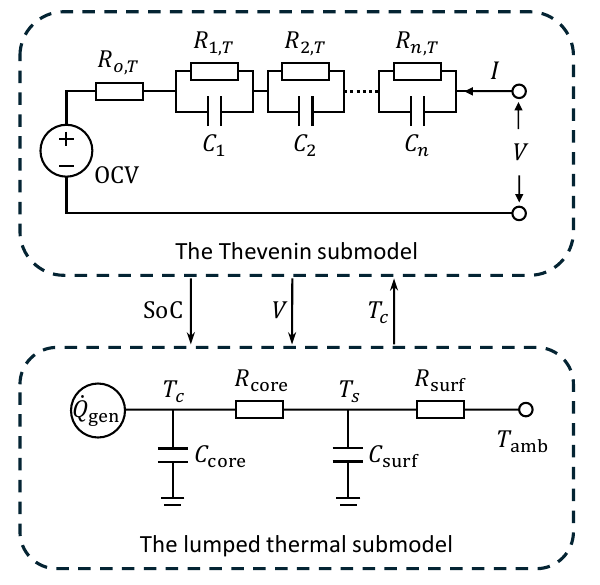}
\caption{The TheveninT model, which couples the Thevenin submodel and the lumped thermal submodel.}
\label{fig:thevenin_schematic}
\end{figure}

The first ECM is the TheveninT model, as shown in Fig.~\ref{fig:thevenin_schematic}. This model couples a general-form Thevenin submodel with a thermal submodel. Here, the polarization voltage across the $i$-th RC pair is given by
\begin{equation*}
\dot{V}_{p,i}(t) = -\frac{1}{R_{i,T}\, C_i} V_{p,i}(t) 
- \frac{1}{C_i} I(t).
\end{equation*}
while the state of charge $\mathrm{SoC}(t)$ evolves according to Coulomb counting. The terminal voltage is expressed as
\begin{equation*}
V(t) = h_{\mathrm{OCV}}(\mathrm{SoC}(t)) - \sum_{i=1}^{n} V_{p,i}(t) + R_{o,T} I(t),
\end{equation*}
where $h_{\mathrm{OCV}}(\cdot)$ denotes the open-circuit voltage, $R_{o,T}$ is the ohmic resistance, and $I(t)$ denotes the input current, with $I(t)<0$ for discharging and $I(t)>0$ for charging.

The thermal behavior is described by a two-node lumped submodel~\cite{LIN:2014:PS}.
\begin{subequations}
\label{eq:Thermal}
\begin{align}
\label{eq:Thermal_Tc}
C_{\mathrm{core}} \dot{T}_c(t) &= \dot Q_{\mathrm{gen}} - \frac{T_c(t) - T_s(t)}{R_{\mathrm{core}}}, \\
\label{eq:Thermal_Ts}
C_{\mathrm{surf}} \dot{T}_s(t) &= \frac{T_c(t) - T_s(t)}{R_{\mathrm{core}}} - \frac{T_s(t) - T_{\mathrm{amb}}}{R_{\mathrm{surf}}}.
\end{align}
\end{subequations}
where $T_c$ represents the core temperature, $T_s$ denotes the surface temperature, and the heat generation rate $\dot Q_{\mathrm{gen}}$ is
\begin{equation*}
\dot Q_{\mathrm{gen}} = I(t)\left(V(t)-h_{\mathrm{OCV}}(\mathrm{SoC}(t))\right).
\end{equation*}
Further, $R_{o,T}$ and $R_{i,T}$ are temperature-dependent
\begin{align*}
R_{o,T} &= R_{o} \exp\!\left(\kappa_{1}\!\left(\frac{1}{T_c}-\frac{1}{T_{\mathrm{ref}}}\right)\right),\\
R_{i,T} &= R_{i} \exp\!\left(\kappa_{2}\!\left(\frac{1}{T_c}-\frac{1}{T_{\mathrm{ref}}}\right)\right), \quad i=1,\dots,n,
\end{align*}
with $T_{\mathrm{ref}}$ the reference temperature and $\kappa_{1},\kappa_{2}$ activation energy parameters.

\begin{figure}[t]
\centering
\includegraphics[width=0.48\textwidth]{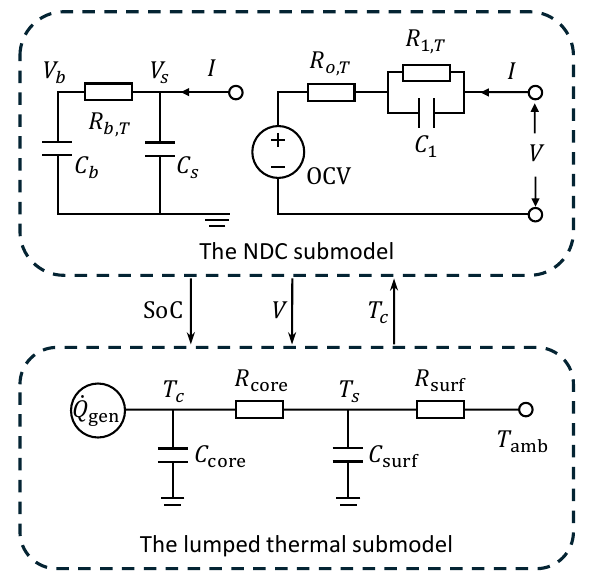}
\caption{The NDCT model, which couples the NDC submodel and the lumped thermal submodel.}
\label{fig:ndct_schematic}
\end{figure}

The second ECM considered here is the NDCT model, which is illustrated in Fig.~\ref{fig:ndct_schematic}. Here, the NDC submodel, developed in~\cite{Tian:2021:TCS}, captures the cell’s electrical dynamics, where the left subcircuit mimics lithium-ion diffusion within the electrode and the right subcircuit characterizes the terminal voltage response. The lumped thermal submodel, in turn, describes the cell’s thermal behavior. The NDC submodel is governed by
\begin{align*}
\dot{V}_b(t) &= \frac{V_s(t) - V_b(t)}{C_b R_{b,T}}, \\
\dot{V}_s(t) &= \frac{V_b(t) - V_s(t)}{C_s R_{b,T}} + \frac{1}{C_s} I(t), \\
\dot{V}_1(t) &= -\frac{V_1(t)}{R_1 C_1} - \frac{1}{C_1} I(t),
\end{align*}
with terminal voltage
\begin{equation*}
V(t) = h_{\mathrm{OCV}}(V_s(t)) - V_1(t) + R_{o,T} I(t).
\end{equation*}
Here, the OCV is dependent on $V_s$, as shown by $h_\mathrm{OCV}(V_s)$. Note that $V_s = \mathrm{SoC}$ when the cell is at equilibrium, i.e., $V_b = V_s$, $V_1 = 0$, and $I=0$.

The thermal dynamics follow the two-node lumped model given in~\eqref{eq:Thermal}.  
The temperature dependence of resistances is described by Arrhenius-type relations:
\begin{align*}
R_{o,T} &= R_{o} \exp\!\left(\kappa_{1}\left(\frac{1}{T_c}-\frac{1}{T_{\mathrm{ref}}}\right)\right), \\
R_{b,T} &= R_{b} \exp\!\left(\kappa_{2}\left(\frac{1}{T_c}-\frac{1}{T_{\mathrm{ref}}}\right)\right).
\end{align*}

\section{System Identification via EnKI} \label{sec:SysID-EnKI}
In this section, we design the system identification approach for the two models introduced in Section~\ref{sec:ECM}. For clarity and consistency, we express both models in a unified state space representation:
\begin{align}
\label{eq:stochastic_ssm}
\begin{aligned}
\dot{\bm{x}}(t) &= f(\bm{x}(t), \bm u(t), \bm{\theta}), \\
\bm y(t) &= g(\bm{x}(t), \bm u(t), \bm{\theta}) + \bm{v}(t),
\end{aligned}
\end{align}
where $\bm{x}(t)$ is the state, $\bm{u}(t)$ the input, $\bm{y}(t)$ the output, $\bm{\theta}$ the parameter vector, and $\bm{v} \sim \mathcal{N}(\bm 0, \bm R)$ the added measurement noise.  More specifically, for the TheveninT model, 
\begin{align*}
\bm{x}(t) &= \begin{bmatrix}
\bm{x}_p & \mathrm{SoC}(t) & T_c(t) & T_s(t)
\end{bmatrix}^\top,\\
\bm \theta &= \big[ R_{o} \quad R_{i} \quad C_{i} \quad C_{\mathrm{core}} \quad C_{\mathrm{surf}} \quad R_{\mathrm{core}} \big. \\
&\quad\quad\quad\quad\quad\quad\quad\quad\quad\quad\quad\quad\quad  \big. R_{\mathrm{surf}} \quad \kappa_{1} \quad \kappa_{2} \big]^\top,
\end{align*}
where $\bm{x}_p(t) = \begin{bmatrix}
V_{p,1}(t) & \cdots & V_{p,n}(t) \end{bmatrix}^\top$, and $i=1,\ldots,n$; for the NDCT model,
\begin{align*}
\bm{x}(t) &= \begin{bmatrix}
V_b(t) & V_s(t) & V_1(t) & T_c(t) & T_s(t)
\end{bmatrix}^\top,\\
\bm \theta &= \big[ C_{b} \quad C_{s} \quad  R_{b} \quad R_{o} \quad C_{\mathrm{core}} \quad C_{\mathrm{surf}} \big. \\
&\quad\quad \quad\quad\quad\quad  \big. R_{\mathrm{core}}\quad R_{\mathrm{surf}} \quad \kappa_{1}\quad \kappa_{2} \quad R_{1} \quad  C_{1} \big]^\top.
\end{align*}
For both models,
\begin{align*}
\bm{u}(t) = \begin{bmatrix} I(t) \\ T_{\mathrm{amb}} \end{bmatrix}, \qquad
\bm{y}(t) = \begin{bmatrix} V(t) \\ T_s(t) \end{bmatrix}.
\end{align*}
In addition, $f(\cdot)$ and $g(\cdot)$ are obvious from each model's context.

Given the state space representation above, we assume that $\bm{u}(t)$ and $\bm{y}(t)$ are measured at discrete times $t_k$ for $k = 1, \ldots, H$, and denote these measurements by $\bm{u}(t_k)$ and $\bm{y}(t_k)$, respectively. The full dataset is then  $\bm D = \left\{ \bm u_1, \ldots, \bm u_H, \bm y_1,  \ldots,  \bm y_H \right\}$. For notational convenience and without loss of generality, we represent the dataset by the outputs alone, $\bm Y := \begin{bmatrix}\bm y_1^{\mathsf T} & \cdots & \bm y_H^{\mathsf T}\end{bmatrix}^{\mathsf T}$. Accordingly, the covariance matrix of the stacked measurement noise is defined as $\bm R_Y:= I_H \otimes \bm R$, where $I_H$ is the identity matrix.To estimate the unknown parameter $\bm{\theta}$ from $\bm{Y}$, we adopt an MAP framework by considering
\begin{align*}
\hat{\bm{\theta}}_{\mathrm{MAP}} = \arg\max_{\bm{\theta}} \p(\bm{\theta} |\bm{Y}).
\end{align*}
However, no closed-form expression is available  for $ \p(\bm{\theta}|\bm{Y})$ as the model in~\eqref{eq:stochastic_ssm} is nonlinear. To address this problem, a straightforward way is approximating $p(\bm \theta, \bm Y)$ by a Gaussian distribution:
\begin{equation}
\label{eq:Joint_Gaussian}
    \begin{bmatrix}
        \bm{\theta} \\
        \bm{Y}
    \end{bmatrix}
    \sim \mathcal{N}
    \left(
    \begin{bmatrix}
        \bar{\bm{\theta}} \\
        \bar{\bm{Y}}
    \end{bmatrix},
    \begin{bmatrix}
        \bm{C}^{\bm{\theta\theta}} & \bm{C}^{\bm{\theta Y}} \\
        (\bm{C}^{\bm{\theta Y}})^\top & \bm{C}^{\bm{YY}} + \bm R_Y
    \end{bmatrix}
    \right),
\end{equation}
where $\bar {\bm \theta}$ and $\bar {\bm Y}$ are the means, and $\bm{C}^{\bm{\theta\theta}}$, $\bm{C}^{\bm Y \bm Y}$, and $\bm{C}^{\bm{\theta Y}}$ are the covariances or cross-covariances. A Gaussian distribution is closed under conditioning, leading to
\begin{align*}
\bm \theta \, | \, \bm Y \sim \mathcal{N} ( \bm m, \bm P ),
\end{align*}
where 
\begin{align*}
\bm m &= \bar{\bm{\theta}} + \bm{C}^{\bm{\theta Y}} \left( \bm{C}^{\bm{YY}} + \bm R_Y \right)^{-1} \left( \bm{Y}   - \bar{\bm{Y}} \right), 
\\
\bm P &=  \bm{C}^{\bm{\theta\theta}} - \bm{C}^{\bm{\theta Y}} \left( \bm{C}^{\bm{YY}} + \bm R_Y \right)^{-1} \left( \bm{C}^{\bm{\theta Y}}  \right)^\top.
\end{align*}
For the approximation in~\eqref{eq:Joint_Gaussian}, we assume that two ensembles of samples, $\left\{ \bm {\theta}^{(1)}, \ldots, \bm{\theta}^{(M)} \right\}$ and  $\left\{ \bm{Y}^{(1)}, \ldots, \bm{Y}^{(M)} \right\}$ form the empirical distributions for $p(\bm \theta)$ and $p(\bm Y)$, respectively. Using them, we can compute the following empirical means and covariances:
\begin{align*}
    \bar{\bm{\theta}} &= \frac{1}{M} \sum_{i=1}^{M} \bm{\theta}^{(i)}, \quad
    \bar{\bm{Y}} = \frac{1}{M} \sum_{i=1}^{M} \bm{Y}^{(i)},
    \\
    \bm{C}^{\bm{\theta\theta}} &= \frac{1}{M - 1} \sum_{i=1}^{M} 
    \left( \bm{\theta}^{(i)} - \bar{\bm{\theta}}  \right)
    \left( \bm{\theta}^{(i)} - \bar{\bm{\theta}}  \right)^\top,
    \\
    \bm{C}^{\bm{\theta Y}} &= \frac{1}{M - 1} \sum_{i=1}^{M} 
    \left( \bm{\theta}^{(i)} - \bar{\bm{\theta}}  \right)
    \left( \bm{Y}^{(i)} - \bar{\bm{Y}}  \right)^\top,
    \\
    \bm{C}^{\bm{YY}} &= \frac{1}{M - 1} \sum_{i=1}^{M} 
    \left( \bm{Y}^{(i)} - \bar{\bm{Y}} \right)
    \left( \bm{Y}^{(i)} - \bar{\bm{Y}} \right)^\top.
\end{align*}

The above derivation gives an ensemble-based update procedure for estimating $\bm\theta$---indeed, $\bm m$ is the estimate of $\bm \theta$ extracted from $\bm Y$, and $\bm P$ is the associated covariance. However, the approximation in~\eqref{eq:Joint_Gaussian} is global in nature and, given the model nonlinearity, lacks accuracy. This, in turn, can significantly compromise the estimation accuracy of $\bm\theta$. To remedy the issue, we impose local Gaussian approximations instead. To show this, we write
\begin{align*}
p(\bm\theta \mid \bm Y) &\propto p(\bm Y \mid \bm\theta)\, p(\bm\theta) \\
&= \left(\prod_{\ell=0}^{L-1} \big[p(\bm Y \mid \bm\theta)\big]^{\alpha_\ell}\right)\, p(\bm\theta),
\end{align*}
where $\alpha_\ell$ is a tempering parameter with $0 < \alpha_\ell < 1$ and $\sum_{\ell=0}^{L-1} \alpha_\ell = 1$, and $\left[\p(\bm{Y} | \bm{\theta})\right]^{\alpha_\ell}$ represents a tempered likelihood. Further, let us define 
\begin{equation}
\label{eq:tempered_posterior}
    p_{\ell+1}(\bm{\theta} \, | \, \bm{Y}) \;\propto\;
    \left[p(\bm{Y} \, | \, \bm{\theta})\right]^{\alpha_\ell}\,
    p_\ell(\bm{\theta} \, | \, \bm{Y}),
\end{equation}
where $p_{\ell}(\bm{\theta} \, | \, \bm{Y})$ is a tempered posterior. Clearly, $p_{\ell}(\bm{\theta} \, | \, \bm{Y}) = p (\bm \theta)$ when $\ell=0$, which is the prior of $\bm \theta$, and $p_{\ell}(\bm{\theta} \, | \, \bm{Y}) = \p(\bm \theta | \bm Y)$ when $\ell = L$. Thus, the relation in~\eqref{eq:tempered_posterior} implies an iterative update of $p_{\ell}(\bm{\theta} \, | \, \bm{Y})$ until convergence to $\p(\bm{\theta} | \bm{Y})$. Note that, if $\bm v \sim \mathcal{N}(\bm 0, \bm R)$, then $\left[\p(\bm{Y} | \bm{\theta})\right]^{\alpha_\ell}$ then has a covariance of $\alpha_\ell^{-1} \bm R_Y$. Now, we impose a local Gaussian approximation  around $\hat {\bm \theta}_\ell$:
\begin{align}
    \begin{bmatrix}
        \bm{\theta} \\
        \bm{Y}
    \end{bmatrix}
    \;\bigg|\;\hat{\bm{\theta}}_\ell
    \;\sim\;
    \mathcal{N}\!\left(
    \begin{bmatrix}
        \hat{\bm{\theta}}_\ell \\
        \hat{\bm{Y}}_\ell
    \end{bmatrix},
    \begin{bmatrix}
        \bm{C}^{\bm{\theta\theta}}_\ell & \bm{C}^{\bm{\theta Y}}_\ell \\
        (\bm{C}^{\bm{\theta Y}}_\ell)^\top & \bm{C}^{\bm{YY}}_\ell + \alpha_\ell^{-1}\bm R_Y
    \end{bmatrix}
    \right).
\end{align}
Hence, $p_{\ell+1}(\bm{\theta} \,|\,\bm{Y}) = \mathcal{N} \left(\bm{\theta}; \hat{\bm \theta}_{\ell+1},  \bm{C}^{\bm{\theta\theta}}_{\ell+1}  \right)$, where
\begin{subequations}
\label{eq:EnKI_Update}
\begin{align}
\hat{\bm{\theta}}_{\ell+1}
&= \hat{\bm{\theta}}_{\ell}
 + \bm{C}^{\bm{\theta Y}}_{\ell}
 \left( \bm{C}^{\bm{YY}}_{\ell} + \alpha_{\ell}^{-1}\bm R_Y \right)^{-1}
 \left( \bm{Y} - \hat{\bm{Y}}_{\ell} \right),
\label{eq:EnKI_mean_update_iter} \\
\bm{C}_{\ell+1} &= \bm{C}^{\bm{\theta\theta}}_\ell - \bm{C}^{\bm{\theta Y}}_\ell \left( \bm{C}^{\bm{YY}}_\ell + \alpha_\ell^{-1}\bm R_Y \right)^{-1} \left( \bm{C}^{\bm{\theta Y}}_\ell \right)^{\top}.
\label{eq:EnKI_cov_update_iter}
\end{align}
\end{subequations}

To implement the procedure in~\eqref{eq:EnKI_mean_update_iter}-\eqref{eq:EnKI_cov_update_iter}, we  leverage ensemble-based empirical distributions for approximation and update the ensembles iteratively as follows:
\begin{subequations}
\label{eq:EnKI_iter}
\begin{align}
    \label{eq:EnKI_update}
    \bm{\theta}_{\ell+1}^{(i)} &= \bm{\theta}_\ell^{(i)} + 
    \bm{C}^{\bm{\theta Y}}_\ell \left( \bm{C}^{\bm{YY}}_\ell + \alpha_\ell^{-1} \bm R_Y \right)^{-1} 
    \left( \bm{Y} - \bm{Y}_\ell^{(i)} \right),\\
    \bar{\bm{\theta}}_\ell &= \frac{1}{M} \sum_{i=1}^{M} \bm{\theta}_\ell^{(i)}, \quad
    \bar{\bm{Y}}_\ell = \frac{1}{M} \sum_{i=1}^{M} \bm{Y}_\ell^{(i)}, \label{eq:EnKI_mean_iter} \\
    \bm{C}^{\bm{\theta\theta}}_\ell &= \frac{1}{M - 1} \sum_{i=1}^{M} 
    \left( \bm{\theta}_\ell^{(i)} - \bar{\bm{\theta}}_\ell \right)
    \left( \bm{\theta}_\ell^{(i)} - \bar{\bm{\theta}}_\ell \right)^\top, \label{eq:EnKI_covth_iter} \\
    \bm{C}^{\bm{\theta Y}}_\ell &= \frac{1}{M - 1} \sum_{i=1}^{M} 
    \left( \bm{\theta}_\ell^{(i)} - \bar{\bm{\theta}}_\ell \right)
    \left( \bm{Y}_\ell^{(i)} - \bar{\bm{Y}}_\ell \right)^\top, \label{eq:EnKI_crosscov_iter}\\
    \bm{C}^{\bm{YY}}_\ell &= \frac{1}{M - 1} \sum_{i=1}^{M} 
    \left( \bm{Y}_\ell^{(i)} - \bar{\bm{Y}}_\ell \right)
    \left( \bm{Y}_\ell^{(i)} - \bar{\bm{Y}}_\ell \right)^\top 
    \label{eq:EnKI_covY_iter}.
\end{align}
\end{subequations}
After completing the $L$ iterations, the final parameter estimate is obtained as the ensemble mean
\begin{equation}
\label{eq:final_theta}
\hat{\bm{\theta}} 
= \frac{1}{M} \sum_{i=1}^{M} \bm{\theta}_{L}^{(i)}.
\end{equation}
The above is the EnKI method, which iteratively updates the ensemble that approximately represents $\p(\bm \theta | \bm Y)$ until achieving convergence. Some further remarks are as follows. 

\begin{rem}
We interpret the EnKI method as resulting from successive local Gaussian approximation for $p\left(\bm \theta, \bm Y\right)$ that yields an iterative procedure. The local approximation overcomes a key issue in the global approximation in~\eqref{eq:Joint_Gaussian}--overconfidence and consequently, strongly biased estimation. This feature thus reduces--though does not eliminate--biases and helps improve the stability in estimation. 
\end{rem}

\begin{rem}
The literature has suggested that the EnKI method enjoys empirical stability~\cite{DUFFIELD:2022:SPL}. For nonlinear models, if the span of the initial ensemble covers the true value of $\bm \theta$, then the subsequent ensembles during the iteration can maintain the coverage, rather than diverge, even for high-dimensional problems. With this, the EnKI method usually converges to meaningful estimates despite certain biases.
\end{rem}

\begin{rem}
Choosing the tempering parameters $\alpha_\ell$ is pivotal. We use the data misfit controller (DMC) proposed in~\cite{Kantas:2014:SIAM}, which adaptively balances the ensemble misfit mean and variance, bounding the information gain at each step and mitigating ensemble collapse. More specifically, for the predicted output ensemble $\{\bm Y_\ell^{(i)}\}_{i=1}^M$ at iteration $\ell$, let $\Phi_\ell^{(i)}$, $\bar{\Phi}_\ell$, and $\sigma_{\Phi_\ell}^2$ denote the data misfit of ensemble member $i$, the empirical mean of the ensemble misfits, and their empirical variance, respectively. Then, $\alpha_\ell$ is selected as
\begin{align*}
\alpha_\ell \;=\; \min\left(
\max\left(\frac{H}{2\bar{\Phi}_\ell},\frac{H}{2\sigma_{\Phi_\ell}^2}\right),
\, 1-t_\ell \right),
\end{align*}
where $t_\ell = \sum_{r=0}^{\ell-1} \alpha_r$ is the cumulative tempering parameter.
\end{rem}

To implement the EnKI method for identifying the state–space model in~\eqref{eq:stochastic_ssm}, we generate an ensemble of $\bm{Y}$ at each iteration. This ensemble serves as an empirical approximation of the distribution $\p(\bm{Y} | \bm{\theta})$. By Bayes’ rule,
\begin{equation*}
\label{eq:Bayes_rule}
\p(\bm{\theta} | \bm{Y}) \;\propto\; \p(\bm{Y} | \bm{\theta}) \, p(\bm{\theta}).
\end{equation*}

Although it is impossible to derive an analytical expression for $p(\bm{Y} \mid \bm{\theta})$, we approximate it using Monte Carlo sampling. Specifically, for a given $\bm{\theta}_\ell^{(i)}$, we propagate the corresponding state trajectory according to
\begin{equation}
\dot{\bm{x}}^{(i)}_{\ell}(t) = f\left(\bm{x}^{(i)}_{\ell}(t), \bm{u}(t), \bm{\theta}^{(i)}_\ell\right).
\label{eq:state}
\end{equation}
At the sampling instants $t_k$,
\begin{equation}
\bm{y}^{(i)}_{\ell}(t_k) = g\left(\bm{x}^{(i)}_{\ell}(t_k), \bm{u}(t_k), \bm{\theta}^{(i)}_\ell\right) + \bm{v}^{(i)}_{\ell}(t_k), 
\label{eq:measurement}
\end{equation}
where $\bm v^{(i)}_{\ell}(t_k)\sim\mathcal N(\bm 0,\bm R)$. This procedure yields the ensemble ${ \bm{Y}_\ell^{(i)} }$, which is subsequently used in~\eqref{eq:EnKI_iter} to facilitate the ensemble update for system identification.

Summarizing the above, Algorithm~\ref{alg:EnKI} shows the proposed EnKI-based system identification approach.
\begin{algorithm}[htbp]
\caption{EnKI-Based System Identification}
\label{alg:EnKI}
\small
\begin{algorithmic}[1]
\State Draw the initial ensemble $\bm{\theta}^{(i)}_{0}$ from a prior $p(\bm\theta)$
\For{$\ell = 0 \; \textbf{to} \; L-1$}
  \For{$i = 1 \; \textbf{to} \; M$}
    \State Generate the ensemble 
      $\bm{Y}_\ell^{(i)}$ using~\eqref{eq:state}-\eqref{eq:measurement}
  \EndFor
  \State Compute $\bar{\bm{\theta}}_\ell$, $\bar{\bm{Y}}_\ell$ via \eqref{eq:EnKI_mean_iter}
  \State Compute $\bm{C}^{\bm{\theta\theta}}_\ell$, $\bm{C}^{\bm{\theta Y}}_\ell$, $\bm{C}^{\bm{YY}}_\ell$ via~\eqref{eq:EnKI_covth_iter}--\eqref{eq:EnKI_covY_iter}
  \State Set tempering parameter $\alpha_\ell$ via the DMC
  \For{$i = 1 \; \textbf{to} \; M$}
    \State Update $\bm{\theta}_{\ell+1}^{(i)}$ using \eqref{eq:EnKI_update}
  \EndFor
\EndFor
\State \textbf{return} $\hat{\bm{\theta}}$ via \eqref{eq:final_theta}
\end{algorithmic}
\end{algorithm}

\section{Numerical Simulation and Results}
\label{sec:Numerical_Simulation}
This section validates the EnKI-based system identification approach for the TheveninT and NDCT models using both simulations and experiments. In the simulation study, we identify each model separately using synthetic data generated from its corresponding nominal model. For experimental validation, we estimate the models from testing data and compare their performance.

\subsection{Simulation Setup and Data Generation}
\label{sec:synthetic_data}
For the simulation study, we consider a cell with a Nickel Cobalt Aluminum Oxide (NCA) cathode. The cell has a rated capacity of 3.3 Ah, a maximum voltage of 4.2 V, and a minimum operating voltage of 2.5 V. Experimental identification of this cell was reported in~\cite{TIAN:2020:ES, BIJU:2023:AE}, and the models obtained therein serve as the nominal models for our study. In the sequel, the nominal model's parameters will serve as the ground truth. We generate synthetic data by applying four different current profiles to the nominal models under various temperature conditions: US06 at 313 K, LA92 at 298 K, UDDS at 283 K, and SC04 at 303 K. In the simulations of both models, the initial states are set to $\text{SoC}(0) = 1$, $V_1(0)=0$, $V_b(0)=V_s(0)=1$, and $T_c(0)=T_s(0)=T_{\mathrm{amb}}$. The data are sampled at $\Delta t = 1~\mathrm{s}$. The current profiles are scaled to the range 0–4 A. The resulting synthetic voltage and temperature measurements are corrupted with additive Gaussian noise, following $\mathcal{N}(0,10^{-4})$ and $\mathcal{N}(0,10^{-3})$, respectively. 
To identify the considered ECMs from these synthetic datasets, we apply the EnKI algorithm with an ensemble size of $M=200$ for a trade-off between the computational cost and the estimation accuracy. The initial EnKI ensemble is drawn from a Gaussian prior $\mathcal{N}(\bm{\mu}_0,\bm{\Sigma}_0)$, where $\bm{\mu}_0=\bm{\theta}_{\mathrm{true}}+0.3\,\operatorname{diag}\,(\bm{\theta}_{\mathrm{true}})\bm{\epsilon}$, $\bm{\epsilon}\sim\mathcal{N}(\mathbf{0},\mathbf{I})$, and $\bm{\Sigma}_0=\operatorname{diag}\,\!\left((0.2\,\bm{\theta}_{\mathrm{true}})^2\right)$.

\begin{figure*}[t]
    \centering
    \vspace{4mm}
    \includegraphics[width=1\linewidth]{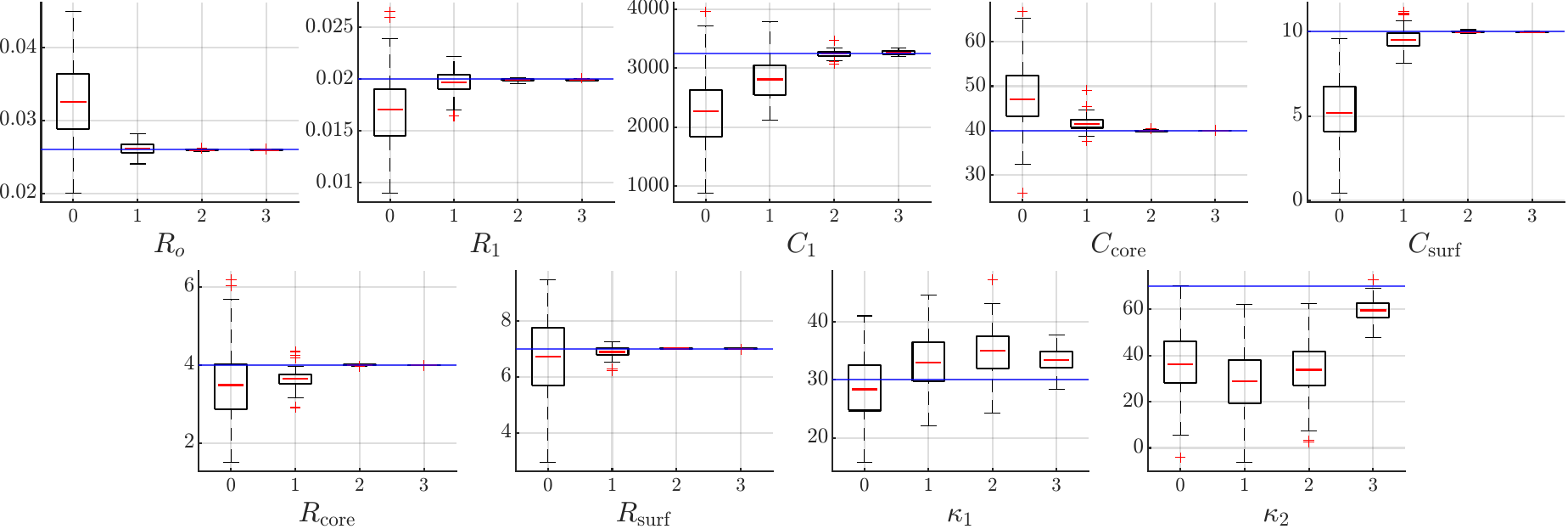}
    \caption{Boxplots of the ensembles for the parameters during the iterations in identifying the TheveninT model. The central red line marks the median, the box spans the interquartile range, the whiskers show the overarching ranges, and the red pluses are the outliers. Horizontal blue lines indicate the true parameter values of the TheveninT model.}
    \label{fig:Thevenin_1RC_boxplot}
\end{figure*}

\begin{figure}[htbp]
    \centering
    \includegraphics[width=0.98\linewidth]{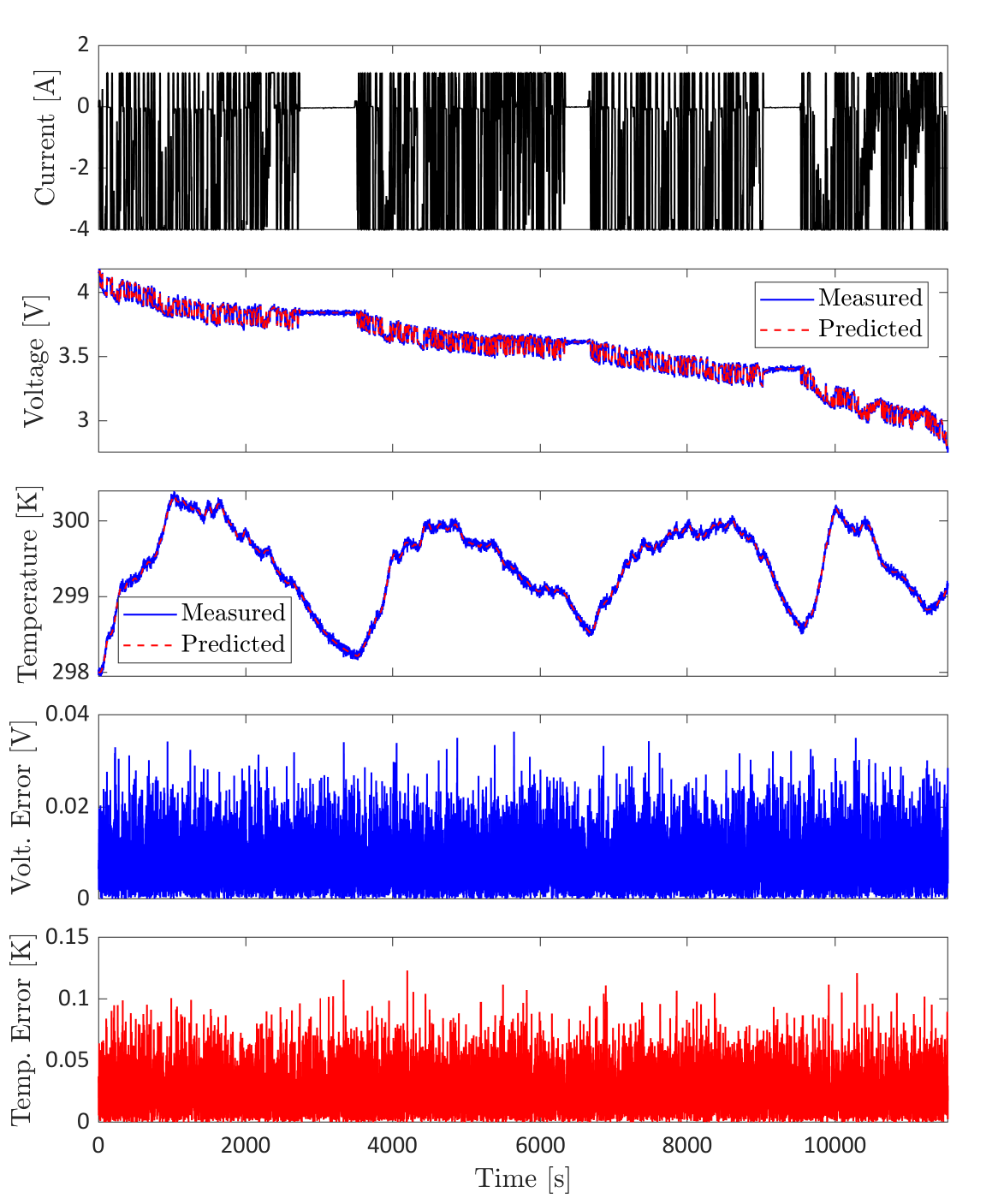}
    \caption{ Comparison of the measured and predicted voltage and surface temperature under the current profile, involving both charging and discharging, at $T_{\mathrm{amb}}=298$ K for the TheveninT model.}
    \label{fig:Thevenin_Comparison}
\end{figure}

Table~\ref{tab:Thevenin_results} summarizes the system identification results for the TheveninT model with one serial RC circuit. Overall, the estimated parameters are close to the nominal values, with most relative errors below 0.3\%. The errors for $\kappa_1$ and $\kappa_2$ are higher, reflecting the relatively low sensitivity of the measurements to these parameters. Fig.~\ref{fig:Thevenin_1RC_boxplot} presents the parameter ensembles as boxplots across iterations. Notably, the EnKI method converges within three iterations, enabled by the choice of $\alpha_\ell$ based on the DMC method. As shown in Fig.~\ref{fig:Thevenin_1RC_boxplot}, the ensembles initially deviate significantly from the nominal values but quickly contract and converge—the medians move toward the nominal values while the spread narrows. This behavior demonstrates the effectiveness of the proposed approach in performing ensemble-based parameter identification. Fig.~\ref{fig:Thevenin_Comparison} further shows the voltage and temperature fitting performance under a composite dynamic current. Both the voltage and temperature predictions of the identified TheveninT model closely track the measured values, consistent with the prediction errors reported in Fig.~\ref{fig:Thevenin_Comparison}.

\begin{table}[t]
\centering
\vspace{2mm}
\caption{Comparison between the true and estimated parameters for the TheveninT model.}
\label{tab:Thevenin_results}
\begin{tabular}{lccc}
\toprule
\textbf{Parameter} & \textbf{True Value} & \textbf{Identified Value} & \textbf{Relative Error (\%)} \\
\midrule
$R_{o}$ [$\Omega$] & 0.026  & 0.02597  &  0.1\\
$R_1$ [$\Omega$] & 0.02  & 0.01993  &  0.31\\
$C_1$ [F] & 3250  & 3254.081 &  0.13\\
$C_{\text{core}}$ [J/K] & 40 & 39.9283 &  0.18\\
$C_{\text{surf}}$ [J/K]  & 10 & 9.9952 &  0.05\\
$R_{\text{core}}$ [K/W]  & 4  & 4.01035 &  0.26\\
$R_{\text{surf}}$ [K/W]  & 7  & 7.00997 &  0.14\\
$\kappa_{1}$ & 30 & 33.3955  &  11.32\\
$\kappa_{2}$ & 70 & 59.5805  &  14.88\\
\bottomrule
\end{tabular}
\end{table}

\begin{figure*}[t]
    \centering
    \vspace{4mm}
    \includegraphics[width=1\linewidth]{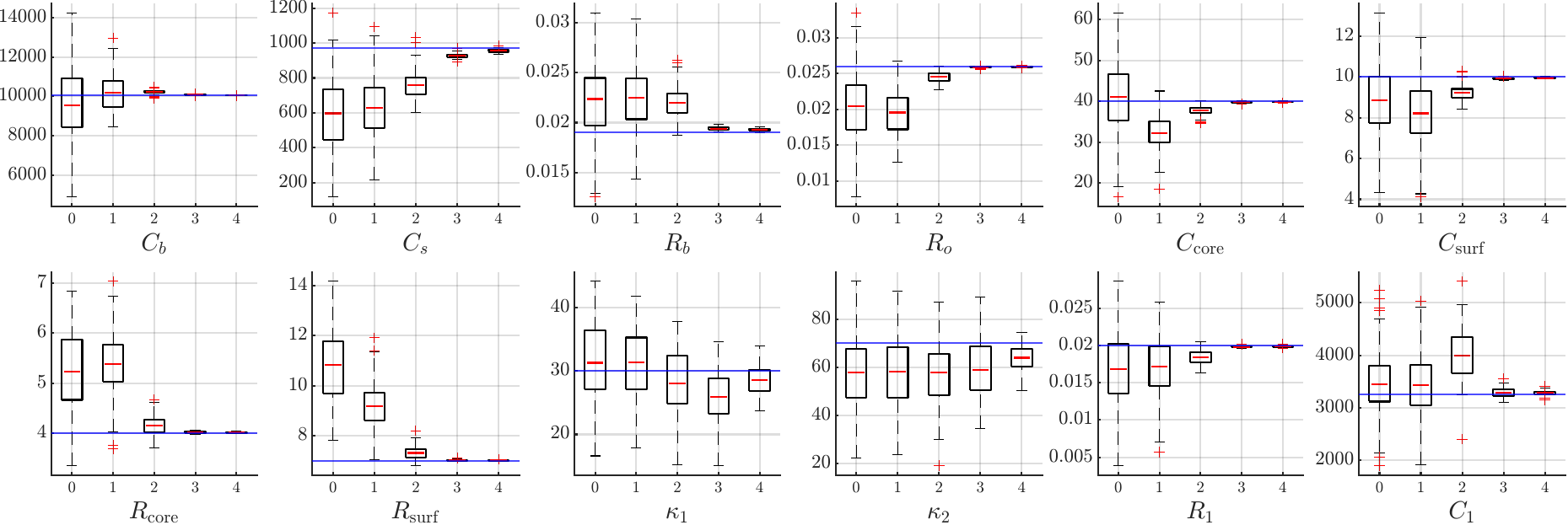}
    \caption{Boxplots of the ensembles for the parameters during the iterations in identifying the NDCT  model. The central red line marks the median, the box spans the interquartile range, the whiskers show the overarching ranges, and the red pluses are the outliers. Horizontal blue lines indicate the true parameter values of the NDCT model.}
    \label{fig:NDCT_Boxplot}
\end{figure*}

\begin{figure}[t]
    \centering
    \includegraphics[width=0.98\linewidth]{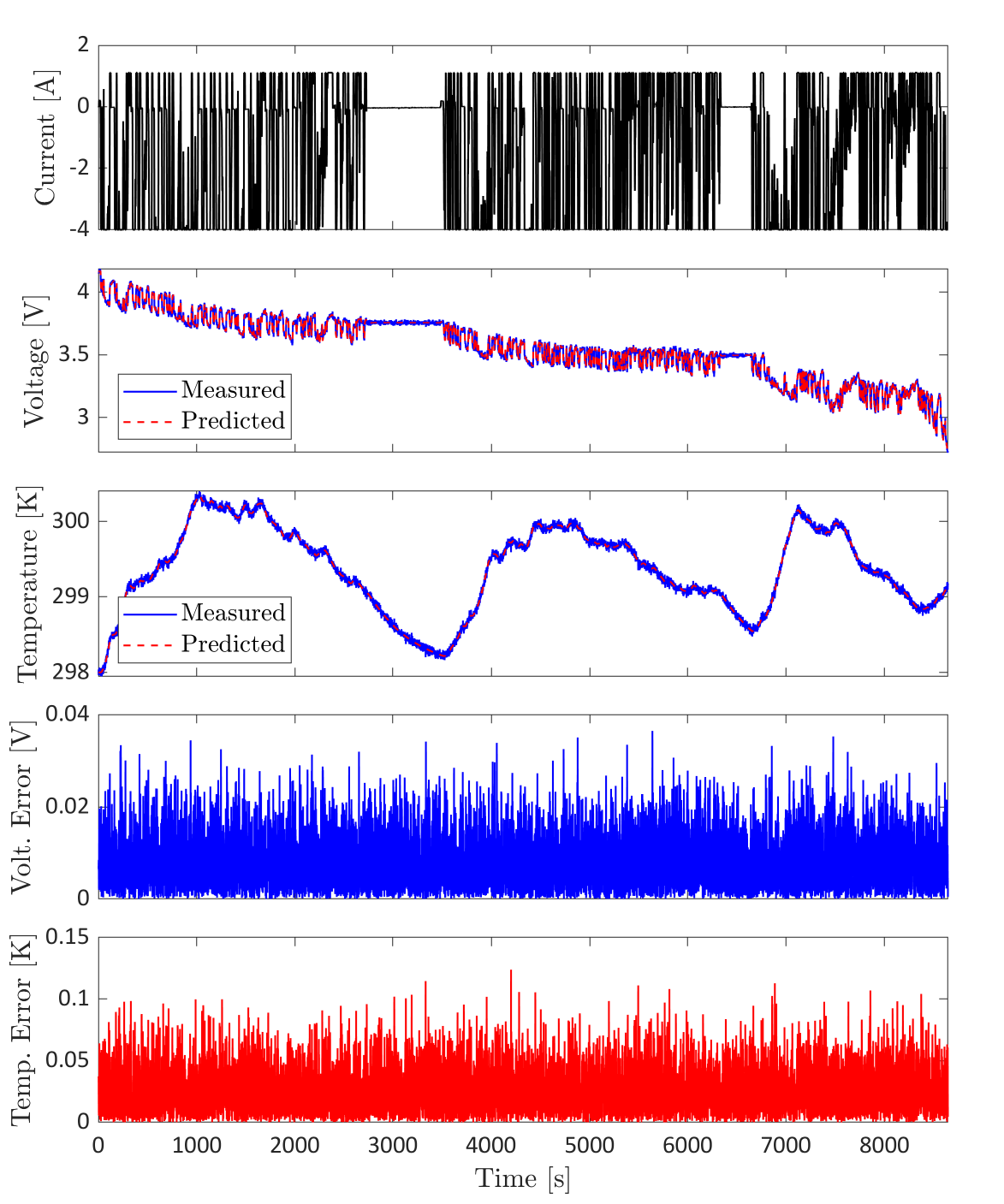}
    \caption{ Comparison of the measured and predicted voltage and surface temperature under the current profile, involving both charging and discharging, at $T_{\mathrm{amb}}=298$ K for the NDCT model.}
    \label{fig:NDCT} 
\end{figure}

We now examine system identification for the NDCT model, which contains 12 parameters—three more than the TheveninT model. Despite this increased complexity, the proposed approach delivers effective performance. As reported in Table~\ref{tab:Identified_params_NDCT}, the estimated parameters closely match the nominal values, with most relative percentage errors below 1\%. The estimates of $\kappa_1$ and $\kappa_2$ are less accurate, again due to the low sensitivity of the measurements to these parameters. Fig.~\ref{fig:NDCT_Boxplot} illustrates the evolution of the parameter ensembles, which concentrate rapidly toward the nominal values within only four iterations. Fig.~\ref{fig:NDCT} further shows the voltage and temperature predictions of the identified NDCT model, which exhibit close agreement with the actual values.

\begin{table}[t]
\centering
\caption{Comparison between the true and estimated parameters for the NDCT model.}
\label{tab:Identified_params_NDCT}
\begin{tabular}{lccc}
\toprule
\textbf{Parameter} & \textbf{True Value} & \textbf{Estimated Value} & \textbf{Relative Error(\%)} \\
\midrule
$C_b$ [F] & 10037 & 10033.38 & 0.04\\
$C_s$ [F] & 973  & 978.96  &  0.61\\
$R_b$ [$\Omega$] & 0.019  & 0.0192 &  1.10\\
$R_o$ [$\Omega$] & 0.026  & 0.02597 &  0.11\\
$C_{\text{core}}$ [J/K] & 40 & 39.9225 & 0.19 \\
$C_{\text{surf}}$ [J/K]  & 10 & 9.9697 &  0.30\\
$R_{\text{core}}$ [K/W]  & 4  & 4.0095 &  0.24\\
$R_{\text{surf}}$ [K/W]  & 7  & 7.0165 &  0.24\\
$\kappa_1$  & 30  & 28.6809 &  4.39\\
$\kappa_2$  & 70 & 64.4315 & 7.95\\
$R_1$ [$\Omega$] & 0.02  & 0.01988  &  0.58\\
$C_1$ [F] & 3250 & 3256.825 &  0.21\\
\bottomrule
\end{tabular}
\end{table}

\subsection{Experimental Validation}
\begin{figure}[t]
    \centering
    \includegraphics[width=0.98\linewidth]{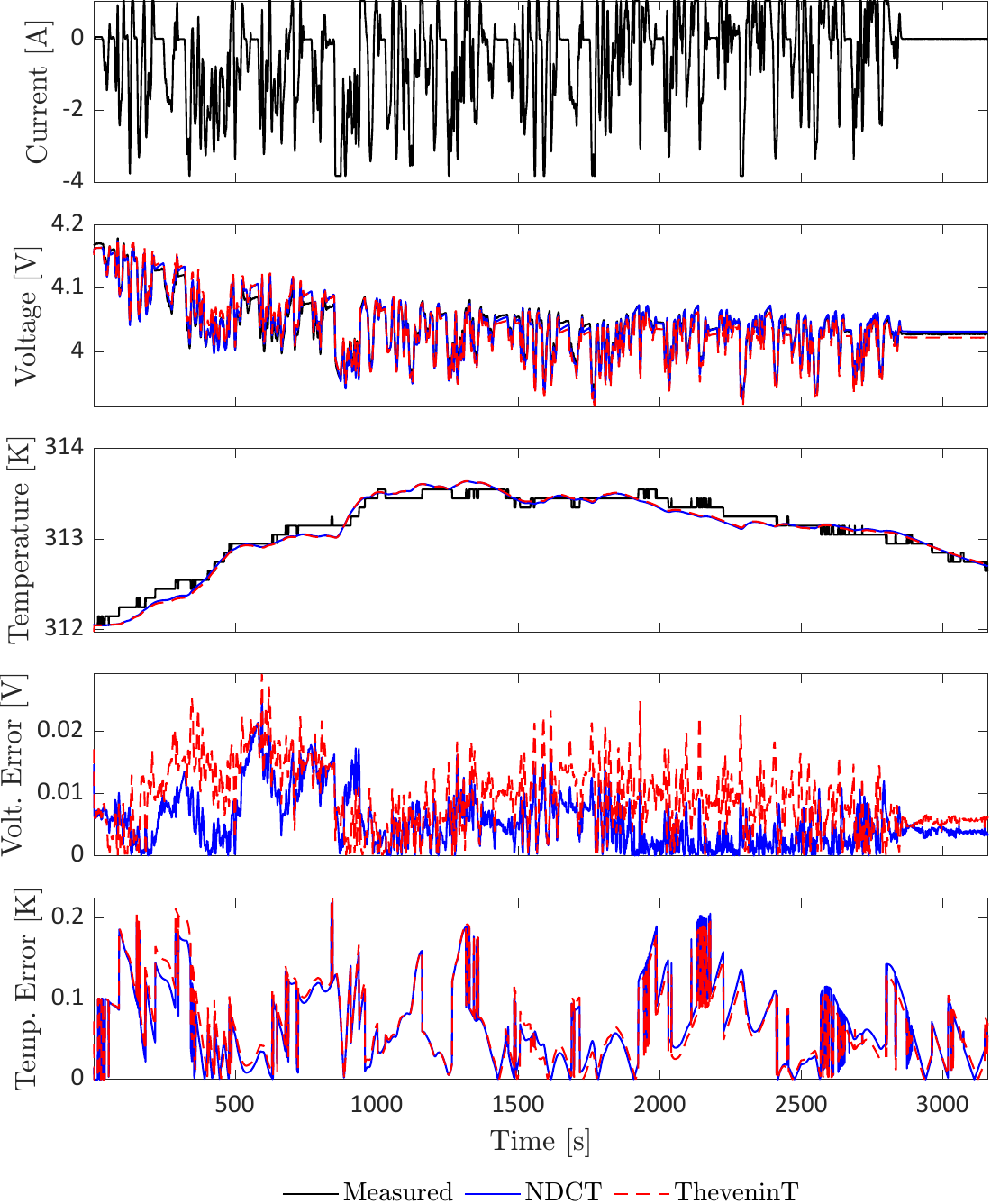}
    \caption{Experimental validation and comparison of the TheveninT and NDCT model in voltage and temperature prediction.}
    \label{fig:Experimental}
\end{figure}

To further evaluate the proposed approach, we collect experimental data from a Samsung INR18650-25R cell using a PECR SBT4050 battery tester. The experiments were carried out at an ambient temperature of $312$ K under the LA92 current profile, spanning from fully charged to completely discharged. We use the identification results from Section~\ref{sec:synthetic_data} as priors to inform this stage of estimation. Fig.~\ref{fig:Experimental} presents the resulting voltage and temperature predictions from the TheveninT and NDCT models. The results demonstrate that the proposed approach enables effective parameter estimation for both models, as their predictions match the measured data. Moreover, the NDCT model achieves noticeably higher accuracy than the TheveninT model, particularly at low and high SoC levels and under high C-rate conditions. This improvement arises because the NDCT model is designed to capture richer lithium-ion battery dynamics, thereby providing better descriptive capability.

\section{Conclusion}
\label{sec:Conclusion}

LiBs have gained ever-growing use to advance electrified transportation and the broader energy transition, with their role expected to expand significantly in the coming decades. Ensuring their safe and efficient operation in practical applications requires accurate dynamic models. Although battery modeling has attracted substantial research interest, existing system identification methods often face limitations in terms of accuracy, applicability to different datasets, and computational tractability. In this paper, we propose an EnKI-based system identification approach for LiBs using state-space models. This method leverages EnKI’s integration of Monte Carlo sampling and Kalman update to perform an iterative, ensemble-based search for model parameters. It is capable of handling strong model nonlinearities, exhibits computational stability, and bypasses the need for derivative computations. We demonstrate the proposed method on two ECMs: the TheveninT model and the NDCT model. Both simulation and experimental results show that the method not only achieves accurate parameter identification, but also converges fast. The approach appears promising for broader application  to other LiB models as well, which we aim to explore in future work.

\FloatBarrier 
\balance
\bibliographystyle{ieeetr}
\bibliography{references.bib}

@article{BIJU:2023:AE,
title = {{BattX}: An equivalent circuit model for lithium-ion batteries over broad current ranges},
journal = {Applied Energy},
volume = {339},
pages = {120905},
year = {2023},
issn = {0306-2619},
author = {Nikhil Biju and Huazhen Fang},
}

@article{Kantas:2014:SIAM,
author = {Kantas, Nikolas and Beskos, Alexandros and Jasra, Ajay},
title = {{Sequential Monte Carlo Methods for High-Dimensional Inverse Problems: A Case Study for the Navier--Stokes Equations}},
journal = {SIAM/ASA Journal on Uncertainty Quantification},
volume = {2},
number = {1},
pages = {464-489},
year = {2014},
}

@article{DUFFIELD:2022:SPL,
title = {Ensemble {Kalman} inversion for general likelihoods},
journal = {Statistics \& Probability Letters},
volume = {187},
pages = {109523},
year = {2022},
issn = {0167-7152},
author = {Samuel Duffield and Sumeetpal S. Singh},
}

@ARTICLE{Tian:2021:TCS,
  author={Tian, Ning and Fang, Huazhen and Chen, Jian and Wang, Yebin},
  journal={IEEE Transactions on Control Systems Technology}, 
  title={Nonlinear Double-Capacitor Model for Rechargeable Batteries: Modeling, Identification, and Validation}, 
  year={2021},
  volume={29},
  number={1},
  pages={370-384},
}

@article{LIN:2014:PS,
title = {A lumped-parameter electro-thermal model for cylindrical batteries},
journal = {Journal of Power Sources},
volume = {257},
pages = {1-11},
year = {2014},
issn = {0378-7753},
author = {Xinfan Lin and Hector E. Perez and Shankar Mohan and Jason B. Siegel and Anna G. Stefanopoulou and Yi Ding and Matthew P. Castanier},
}

@article{Iglesias:2018:IP,
year = {2018},
publisher = {IOP Publishing},
volume = {34},
number = {10},
pages = {105002},
author = {Iglesias, Marco and Park, Minho and Tretyakov, M V},
title = {{Bayesian} inversion in resin transfer molding},
journal = {Inverse Problems},
}

@article{Kovachki:2019:IP,
year = {2019},
publisher = {IOP Publishing},
volume = {35},
number = {9},
pages = {095005},
author = {Kovachki, Nikola B and Stuart, Andrew M},
title = {Ensemble {Kalman} inversion: a derivative-free technique for machine learning tasks},
journal = {Inverse Problems},
}

@article{Iglesias:2013:IP,
year = {2013},
publisher = {IOP Publishing},
volume = {29},
number = {4},
pages = {045001},
author = {Iglesias, Marco A and Law, Kody J H and Stuart, Andrew M},
title = {Ensemble {Kalman} methods for inverse problems},
journal = {Inverse Problems},
}

@INPROCEEDINGS{Tu:2024:ACC,
  author={Tu, Hao and Lin, Xinfan and Wang, Yebin and Fang, Huazhen},
  booktitle={Proceedings of the American Control Conference},  
  title={System Identification for Lithium-Ion Batteries with Nonlinear Coupled Electro-Thermal Dynamics via {Bayesian} Optimization}, 
  year={2024},
  volume={},
  number={},
  pages={1946-1951},
}

@article{YANG:2014:PS,
title = {Battery parameterisation based on differential evolution via a boundary evolution strategy},
journal = {Journal of Power Sources},
volume = {245},
pages = {583-593},
year = {2014},
issn = {0378-7753},
author = {Guangya Yang},
}

@ARTICLE{YU:2017:IEEE,
  author={Yu, Zhihao and Xiao, Linjing and Li, Hongyu and Zhu, Xuli and Huai, Ruituo},
  journal={IEEE Transactions on Industrial Electronics}, 
  title={Model Parameter Identification for Lithium Batteries Using the Coevolutionary Particle Swarm Optimization Method}, 
  year={2017},
  volume={64},
  number={7},
  pages={5690-5700},
}

@article{MALIK:2014:PS,
title = {Extraction of battery parameters using a multi-objective genetic algorithm with a non-linear circuit model},
journal = {Journal of Power Sources},
volume = {259},
pages = {76-86},
year = {2014},
issn = {0378-7753},
author = {Aimun Malik and Zheming Zhang and Ramesh K. Agarwal},
}

@Article{Samieian:2022:B,
AUTHOR = {Samieian, Mohammad Amin and Hales, Alastair and Patel, Yatish},
TITLE = {A Novel Experimental Technique for Use in Fast Parameterisation of Equivalent Circuit Models for Lithium-Ion Batteries},
JOURNAL = {Batteries},
VOLUME = {8},
YEAR = {2022},
NUMBER = {9},
ARTICLE-NUMBER = {125},
ISSN = {2313-0105},
}

@ARTICLE{Sitterly:2011:IEEE,
  author={Sitterly, Mark and Wang, Le Yi and Yin, G. George and Wang, Caisheng},
  journal={IEEE Transactions on Sustainable Energy}, 
  title={Enhanced Identification of Battery Models for Real-Time Battery Management}, 
  year={2011},
  volume={2},
  number={3},
  pages={300-308},
}

@article{FENG:2015:PS,
title = {Online identification of lithium-ion battery parameters based on an improved equivalent-circuit model and its implementation on battery state-of-power prediction},
journal = {Journal of Power Sources},
volume = {281},
pages = {192-203},
year = {2015},
issn = {0378-7753},
author = {Tianheng Feng and Lin Yang and Xiaowei Zhao and Huidong Zhang and Jiaxi Qiang},
}

@article{LUDWIG:2021:PS,
title = {Pulse resistance based online temperature estimation for lithium-ion cells},
journal = {Journal of Power Sources},
volume = {490},
pages = {229523},
year = {2021},
issn = {0378-7753},
author = {S. Ludwig and I. Zilberman and M.F. Horsche and T. Wohlers and A. Jossen},
}

@Article{Zhang:2020:en,
AUTHOR = {Zhang, Liang and Wang, Shunli and Stroe, Daniel-Ioan and Zou, Chuanyun and Fernandez, Carlos and Yu, Chunmei},
TITLE = {An Accurate Time Constant Parameter Determination Method for the Varying Condition Equivalent Circuit Model of Lithium Batteries},
JOURNAL = {Energies},
VOLUME = {13},
YEAR = {2020},
NUMBER = {8},
ARTICLE-NUMBER = {2057},
ISSN = {1996-1073},
}

@ARTICLE{Chen:2006:IEEE,
  author={Chen, Min and Rincon-Mora, G.A.},
  journal={IEEE Transactions on Energy Conversion}, 
  title={Accurate electrical battery model capable of predicting runtime and {I-V} performance}, 
  year={2006},
  volume={21},
  number={2},
  pages={504-511},
}

@article{TIAN:2020:ES,
title = {One-shot parameter identification of the {Thevenin’s} model for batteries: Methods and validation},
journal = {Journal of Energy Storage},
volume = {29},
pages = {101282},
year = {2020},
issn = {2352-152X},
author = {Ning Tian and Yebin Wang and Jian Chen and Huazhen Fang},
}

@article{YUAN:2025:ATE,
title = {A thermal modeling approach for pouch lithium-ion batteries based on mode-thermal power mapping},
journal = {Applied Thermal Engineering},
volume = {280},
pages = {128296},
year = {2025},
issn = {1359-4311},
author = {Qingyang Yuan and Zhonggui Zhang and Bo Zhang},
}

@article{GENG:2021:EA,
title = {Bridging physics-based and equivalent circuit models for lithium-ion batteries},
journal = {Electrochimica Acta},
volume = {372},
pages = {137829},
year = {2021},
issn = {0013-4686},
author = {Zeyang Geng and Siyang Wang and Matthew J. Lacey and Daniel Brandell and Torbjörn Thiringer},
}

@article{DeOliveira:Heliyon:2024,
title = {A two-step identification approach for an extended nonlinear double-capacitor model},
journal = {Heliyon},
volume = {10},
number = {18},
pages = {e36845},
year = {2024},
issn = {2405-8440},
author = {Jose {Genario de Oliveira} and Cisel Aras and Pankaj Pallewar and Mohammad Charkhgard and Thyagesh Sivaraman and Christoph Hametner},
}

@article{MOUSAVIG:2014:RSE,
title = {Various battery models for various simulation studies and applications},
journal = {Renewable and Sustainable Energy Reviews},
volume = {32},
pages = {477-485},
year = {2014},
issn = {1364-0321},
author = {S.M. {Mousavi G.} and M. Nikdel},
}

@Article{He:2011:E,
AUTHOR = {He, Hongwen and Xiong, Rui and Fan, Jinxin},
TITLE = {Evaluation of Lithium-Ion Battery Equivalent Circuit Models for State of Charge Estimation by an Experimental Approach},
JOURNAL = {Energies},
VOLUME = {4},
YEAR = {2011},
NUMBER = {4},
PAGES = {582--598},
ISSN = {1996-1073},
}

@article{HU:2012:PS,
title = {A comparative study of equivalent circuit models for Li-ion batteries},
journal = {Journal of Power Sources},
volume = {198},
pages = {359-367},
year = {2012},
issn = {0378-7753},
author = {Xiaosong Hu and Shengbo Li and Huei Peng},
}

@book{plett:2015:bms,
  author = {Gregory L. Plett},
  title = {Battery Management Systems: Battery Modeling},
  volume = {1},
  publisher = {Artech House},
  address = {Norwood, MA, USA},
  year = {2015}
}

@article{KAPOOR:2025:MSE,
title = {Insights into advances in flexible lithium-ion battery energy storage systems toward sustainable applications},
journal = {Materials Science and Engineering: B},
volume = {318},
pages = {118301},
year = {2025},
issn = {0921-5107},
author = {Ashish Kapoor and Amit Kumar Rathoure and M. Monica and Asmita Chakraborty and S.C. Tripathi and Dan Bahadur Pal},
}

@article{NGOY:2025:RSE,
title = {Lithium-ion batteries and the future of sustainable energy: A comprehensive review},
journal = {Renewable and Sustainable Energy Reviews},
volume = {223},
pages = {115971},
year = {2025},
issn = {1364-0321},
author = {Kitalu Ricin Ngoy and Valantine Takwa Lukong and Kelvin O. Yoro and John Beya Makambo and Nonso Christopher Chukwuati and Chinedu Ibegbulam and Orevaoghene Eterigho-Ikelegbe and Kingsley Ukoba and Tien-Chien Jen},
}

\end{document}